\documentclass{elsart3p}

 \usepackage{epsfig}

\usepackage{amssymb}

\begin{document}

\begin{frontmatter}



\title{Improved model for the analysis of air fluorescence induced by electrons}


\author{F. Arqueros, F. Blanco and J. Rosado}

\address{
Facultad de Ciencias F\'{\i}sicas, Universidad Complutense de Madrid, E-28040 Madrid, Spain}

\begin{abstract}
A model recently proposed for the calculation of air-fluorescence yield excited by electrons is revisited. Improved energy
distributions of secondary electrons and a more realistic Monte Carlo simulation including some additional processes have
allowed us to obtain more accurate results. The model is used to study in detail the relationship between fluorescence intensity
and deposited energy in a wide range of primary energy (keVs - GeVs). In addition, predictions on the absolute value of the
fluorescence efficiency in the absence of collisional quenching will be presented and compared with available experimental data.
\end{abstract}

\journal{5th Fluorescence Workshop, Madrid, 2007}
\end{frontmatter}

\section{Introduction}
\label{intro} The  accuracy in the energy determination of ultra-high energy cosmic rays using the fluorescence technique is
presently limited by the uncertainty in the air-fluorescence yield. In order to improve the accuracy of this fundamental
parameter, several laboratory measurements are being performed \cite{nagano2,flash,flash2,colin,airfly1,paris,tilo,rosado}. On
the other hand, progress on the theoretical understanding of the various processes leading to the air-fluorescence emission is
being carried out. At high pressure ($\gtrsim$ 1 Torr) and high electron energy ($\gtrsim$ 1 keV) most of the fluorescence light
is generated by low energy secondary electrons arising from N$_2$ ionization \cite{blanco}. A simple model which accounts for
the contribution of secondary electrons to the fluorescence yield of the N$_2$ 2P system (C$^3\Pi_u$ $\rightarrow$ B$^3\Pi_g$)
has been recently published \cite{blanco}. Using the above model, the total fluorescence yield has been computed for the first
time at high electron energies in \cite{arqueros}.

\par

Since fluorescence is mainly generated by secondary electrons, their energy distribution is a key parameter. Previous results
\cite{blanco,arqueros} were obtained using the analytical energy distribution given by Opal et al. \cite{opal} which has been
experimentally checked at low energy (below a few keVs). Unfortunately, the extrapolation of this formula to higher energies is
not suitable to account for the energy distribution of delta rays properly. In the present paper, a new analytical approach
valid in the whole energy interval (eV - GeV) is used. Preliminary updated results of our model using this energy spectrum
together with improved molecular parameters and a more realistic Monte Carlo simulation are presented in this paper.

\par

Proportionality between fluorescence intensity and the energy deposited by the primary electron is usually assumed. Experimental
results seem to support this assumption, at least, in small energy intervals. Recently the MACFLY collaboration has extended
this experimental test to a large energy range, 1.5 MeV - 50 GeV, finding proportionality within their experimental errors of
about 15\% \cite{colin}.

\par

However, from a theoretical point of view this proportionality is not obvious. The excitation of the fluorescence emission is
strongly peaked at low energies, in particular for the 2P system of nitrogen and therefore the ratio of fluorescence emission
and deposited energy is strongly dependent on the spectrum of low energy secondaries. In principle, this spectrum depends on
both the primary energy and the distance from the primary interaction, and thus, this proportionality has to be demonstrated
with a detailed analysis.

\section{The model}
\label{MO}

The model described in \cite{blanco,arqueros} shows that the total number of photons of the $v-v'$ band generated by an electron
per unit path length in a thin target can be expressed by

\small
\begin{equation}
\label{FY1} \varepsilon _{vv'}(P) = N \frac{1}{1+P/P'_v}\: \{\sigma_{vv'} (E) + \alpha_{vv'}(E,P)\sigma_{ion}(E)
 \}\,.
\end{equation}
\normalsize

\par

$N$ is the number of nitrogen molecules per unit volume. Collisional de-excitation (quenching) is taken into account by the
well-known Stern-Volmer factor which depends on the characteristic pressure $P'_v$. As shown in this expression, the
fluorescence intensity is a sum of two contributions: a) direct excitation of $v-v'$ band by the primary electron, given by the
optical cross section $\sigma_{vv'}$ and b) contributions from secondary electrons arising from molecular ionizations,
proportional to the ionization cross section.

\par

The optical cross section is defined by

\begin{equation}
\label{PY} \sigma_{vv'} = \sigma_v\frac{A_{vv'}}{\sum_{v'}{A_{vv'}}}\,,
\end{equation}
where $\sigma_{v}$ is the cross section for electron excitation (ionization) of molecular nitrogen in the ground state to the
upper $v$ level of an excited electronic state of N$_2$ (N$_2^+$), and $A_{vv'}$ are the $v-v'$ radiative transition
probabilities (Einstein coefficients). Therefore the optical cross sections $\sigma_{2P(0-0)}$, $\sigma_{1N(0-0)}$ used below
are associated to the emission of a 337 nm and 391 nm photon respectively induced by electron collision on a N$_2$ molecule.

\par

For high pressures ($\gtrsim$ 1 Torr) and energies \linebreak ($\gtrsim$ 1 keV) the contribution from secondary electrons is
dominant for generation of 2P fluorescence and relevant for the 1N bands.
The parameter $\alpha_{vv'}$ in (\ref{FY1}) represents the average number of $v-v'$ photons generated inside the observation
volume as a result of a primary ionization in the absence of quenching.

\par

The evaluation of $\alpha_{vv'}$ has to be carried out by a Monte Carlo simulation which takes into account the various
generations of secondary electrons produced after a primary ionization. The $\alpha_{vv'}$ parameter turns out to be a function
of primary energy $E$, pressure $P$ and the geometrical features of the observation region.

\begin{figure}[t]
\begin{center}
\includegraphics*[width=0.47\textwidth,angle=0,clip]{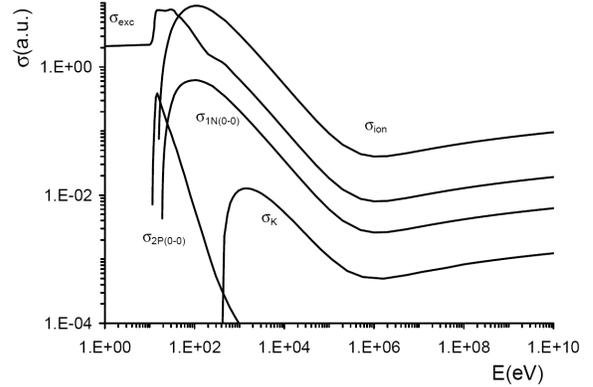}
\caption{\label {fig1} Cross sections (atomic units) involved in the emission of air fluorescence as a function of energy. Cross
section $\sigma_{exc}$ accounts for all excitation processes without emission of a secondary electron. Cross section
$\sigma_{ion}$ accounts for all processes leading to the emission of a secondary electron. The cross section for the ejection of
a K shell electron $\sigma_K$ is shown for comparison. The optical cross sections for the excitation of the 1N (0-0) and 2P
(0-0) bands are also shown.}
\end{center}
\end{figure}

\par

Next we will describe the molecular parameters used in our simulation. In the first place, figure 1 shows the nitrogen cross
section versus energy for various e$^-$-molecule interaction processes of interest in our problem. The cross section for
molecular excitation $\sigma_{exc}$ accounts for all excitation processes without emission of a secondary electron. On the other
hand, the ionization cross section $\sigma_{ion}$ accounts for all processes leading to the emission of a secondary electron and
can be described at large energy by the Born-Bethe function. At high energy only optically allowed transitions, also well
described by a Born-Bethe function, contribute and therefore $\sigma_{exc}$ is proportional to $\sigma_{ion}$. The cross section
for the ejection of a K shell electron $\sigma_K$ is shown for comparison. Figure 1 also shows the optical cross sections
(\ref{PY}) for the excitation of the 2P (0-0) and 1N (0-0) bands of N$_{2}$ and N$_{2}^{+}$ respectively. At low energy,
experimental values have been used (see \cite{arqueros} for references) and at high energies $\sigma_{1N(0-0)}$ has been
extrapolated using the Born-Bethe function. For all cross sections the density correction at high energies
\cite{arqueros,seltzer,perkins} has been applied.

\par

Another ingredient of the model is the energy lost by the electrons in each collision process. For molecular excitation (without
electron ejection) an average value of $<E_{exc}>$ = 8.5 eV has been estimated. In an ionization process the lost kinetic energy
(primary and secondary electrons) equals the N$_2$ ionization potential, $I$ = 15.5 eV, plus the excitation energy of the ion,
for which an average value of \linebreak $<E_{exc}^{ion}>$ = 1.3 eV has been estimated at low energies from available data. At
large energy K-shell ionization is also taken into account with a relative probability given by the corresponding cross section.
In this case the kinetic energy is reduced by 410 eV (K-shell binding energy). As a result $<E_{exc}^{ion}>$ increases up to 5
eV.

\par

In the present calculations the tracks of secondary electrons have been simulated using available angular cross section for the
various individual interaction processes, including elastic e$^-$-molecule collisions. Other processes like generation of X ray
from molecular de-excitation and bremsstrahlung by high energy electrons have also been included in our simulation although the
final results are nearly independent on these processes.

\begin{figure}[t]
\begin{center}
\includegraphics*[width=0.47\textwidth,angle=0,clip]{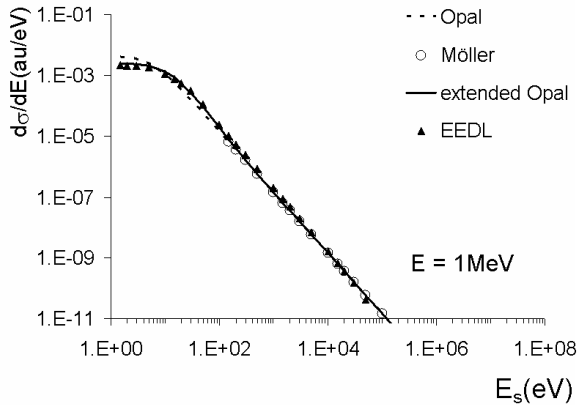}
\caption{\label {fig2} Energy spectrum of secondary electrons emitted in a primary ionization induced by an electron of energy 1
MeV. The Opal formula is not able to reproduce the energy spectrum of delta rays (M\"oller scattering). The extended Opal
formula used in this work fits properly the M\"oller spectrum. The values of the EEDL database are also shown for comparison. In
ordinates the absolute value of the differential cross section in atomic units per eV is represented.}
\end{center}
\end{figure}

\begin{figure}[t]
\begin{center}
\includegraphics*[width=0.47\textwidth,angle=0,clip]{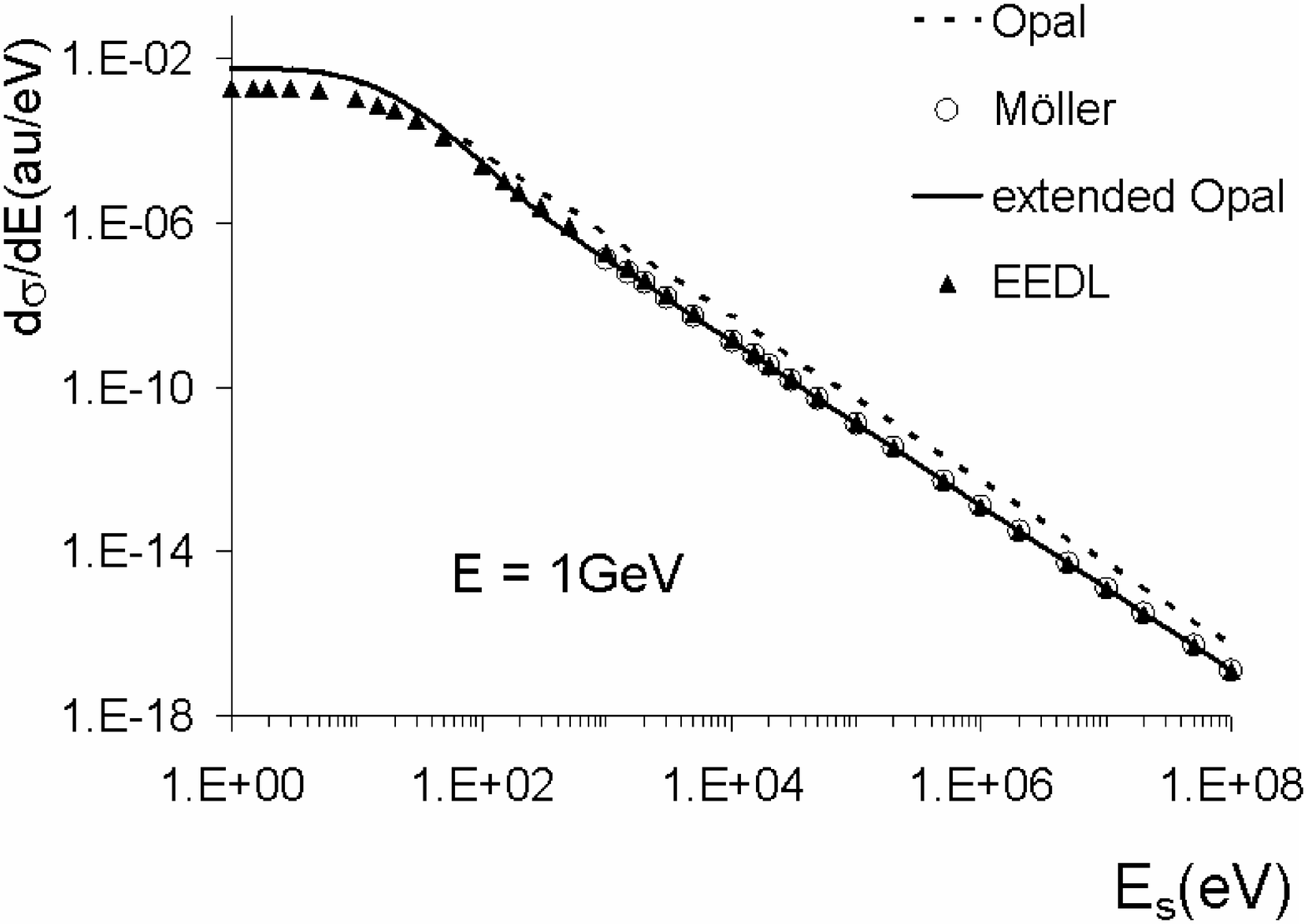}
\caption{\label {fig3} Same as figure 2 for 1 GeV primary energy.}
\end{center}
\end{figure}

As already mentioned, the total fluorescence emission is very sensitive to the energy spectrum of secondary electrons. At low
energy the analytical expression proposed by Opal et al. fits accurately experimental results. At high primary energies, for
which no experimental data are published, precise calculations are described by Seltzer \cite{seltzer,perkins}. Numerical
results for any atom are available in EEDL databases \cite{perkins,EEDL}. Instead of implementing these databases in our
simulation, we looked for an analytical expression for the energy spectrum in the whole energy range for both primary and
secondary electrons. The desired function must fulfill several requirements: \linebreak a) it has to behave as the Opal formula
at low primary energy; b) at high $E$ the function has to account for the well known energy spectrum of delta rays, c) the
integral of the differential cross section has to give the total ionization cross section and d) the assumed function must be
consistent with the collision energy loss given by the Bethe-Bloch theory for electrons.

\par

In this work a new analytical formula has been used which properly accounts for the above four conditions. Details on this
analytical approach will be given in a paper in preparation. In figures 2 and 3 the energy spectrum of secondary electrons
predicted by this ``extended" Opal formula used in the work is compared with other available data for primary electrons of 1 MeV
and 1 GeV respectively.

\par

Fulfilment of requirement d) can be easily checked as follows. The energy loss per unit electron path length in air, assuming
$N_{air}$ molecules per unit volume, is closely related with the mean value of the energy distribution of secondary electrons by
the following relation

\small
\begin{equation}
\label{energy-loss} \frac{{\rm d}E}{{\rm d}x}= N_{air}\{<E_{dep}^0> + <E_{s}>\} \sigma_{ion}(E)\,,
\end{equation}
\normalsize where $<E_{s}>$ is the mean value of the energy distribution of secondary electrons at a given $E$ value and

\small
\begin{equation}
\label{edep_0} <E_{dep}^0> = <E_{exc}>\frac{\sigma_{exc}}{\sigma_{ion}} + I + <E_{exc}^{ion}>\,,
\end{equation}
\normalsize represents the average value of the energy deposited in the medium by the primary electron per primary ionization
process.

\par

In Figure 4 the energy loss calculated from expression (\ref{energy-loss}) using the $<E_{s}>$ value from our energy
distribution of secondary electrons is represented, showing full agreement with the Bethe-Bloch formula \cite{bethe-bloch1}.

\begin{figure}[t]
\begin{center}
\includegraphics*[width=0.47\textwidth,angle=0,clip]{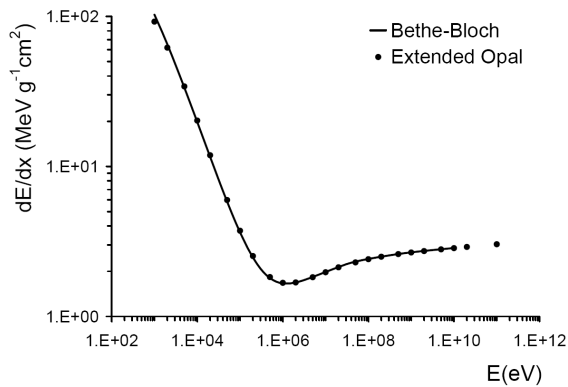}
\caption{\label {fig4} Energy loss of electrons per unit path length versus primary energy. Continuous line represents the
calculated values from equation (\ref{energy-loss}) using the average energy of the extended Opal formula. Black dots are the
values from the Bethe-Bloch formula. In both cases the medium is air at atmospheric pressure.}
\end{center}
\end{figure}

\section{Results}
In this section we will present results on the various parameters of interest for the interpretation of the air fluorescence
measurements. In our simulation the observation region has been assumed to be a sphere of radius $R$ with center in the
interaction point of the primary electron. Assuming a fixed temperature, both fluorescence intensity and deposited energy turn
out to be basically dependent on the product $PR$, that is, on the number of interactions inside the observation region. In
fact, due to the density correction at very large energies, for a given PR value, the above parameters are slightly dependent on
pressure. However, the corresponding deviation is not significant inside the range 100 - 760 Torr.

\begin{figure}[t]
\begin{center}
\includegraphics*[width=0.47\textwidth,angle=0,clip]{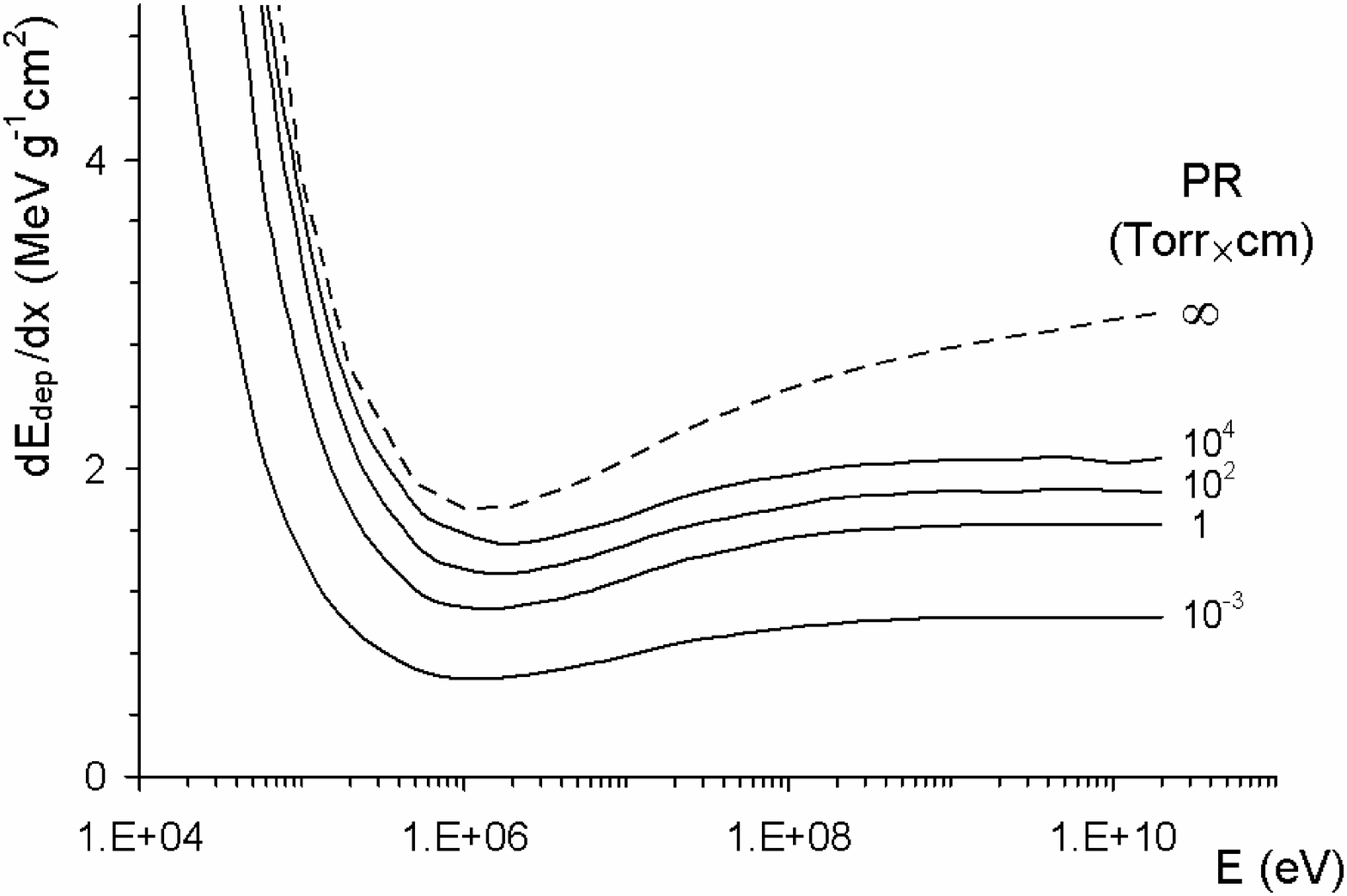}
\caption{\label {fig5} Energy deposited by a primary electron per unit path length versus primary energy for several values of
$PR$ (continuous lines). For very high $PR$ value deposited energy equals the energy loss of the primary electron predicted by
the Bethe-Bloch theory (broken line).}
\end{center}
\end{figure}

The energy deposited by the primary electron per unit path length can be calculated from the average energy deposited by
secondary electrons $<E_{dep}>$ as

\small
\begin{equation}
\label{energy-dep} \frac{{\rm d}E_{dep}}{{\rm d}x}= N_{air}\{<E_{dep}^0> + <E_{dep}>\} \sigma_{ion}(E)\,.
\end{equation}
\normalsize

Note that this equation is the same as (\ref{energy-loss}) replacing $<E_s>$ by $<E_{dep}>$. The value of $<E_{dep}>$ has been
obtained from our Monte Carlo simulation. Figure 5 shows the energy deposited per unit path length versus primary energy for
several $PR$ values. We have checked that, as expected, the deposited energy for an unlimited medium, $PR \rightarrow \infty$,
equals the energy loss predicted by the Bethe-Bloch theory.

\par

The fluorescence efficiency $\Phi_{vv'}$, defined as the ratio between the number of fluorescence photons emitted in a given
$v-v'$ molecular band and the deposited energy is a fundamental parameter. Combining equations (\ref{FY1}) and
(\ref{energy-dep}), the following useful relation can be obtained:

\small
\begin{equation}
\label{Phi0} \Phi_{vv'}^0 = \frac{N}{N_{air}}\times
\frac{\frac{\sigma_{vv'}}{\sigma_{ion}}+\alpha_{vv'}}{<E_{dep}^0>+<E_{dep}>}\,,
\end{equation}
\normalsize
where $\Phi_{vv'}^0$ is the fluorescence efficiency in the absence of collisional quenching and thus related with
$\Phi_{vv'}$ by

\small
\begin{equation}
\label{Phi1} \Phi_{vv'} = \Phi_{vv'}^0 \frac{1}{1+P/P'_v}\,.
\end{equation}
\normalsize

The values of $\alpha_{vv'}$ and $\Phi^0_{vv'}$ have been computed for the 2P (0-0) and 1N (0-0) bands, for which accurate
experimental results of the corresponding cross sections are available. Figures 6 and 7 show $\Phi^0$ against $E$ for several
values of $PR$ while in figures 8 and 9 the dependence of $\Phi^0$ versus $PR$ is represented for various energies.

\begin{figure}[t]
\begin{center}
\includegraphics*[width=0.47\textwidth,angle=0,clip]{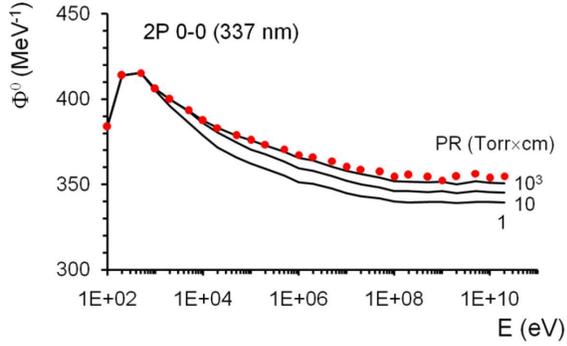}
\caption{\label {fig6} Fluorescence efficiency for the 2P (0-0) band in the absence of collisional quenching against primary
energy for several values of $PR$ (continuous lines). Circles represent the result for a very large medium.}
\end{center}
\end{figure}

\par
\begin{figure}[t]
\begin{center}
\includegraphics*[width=0.47\textwidth,angle=0,clip]{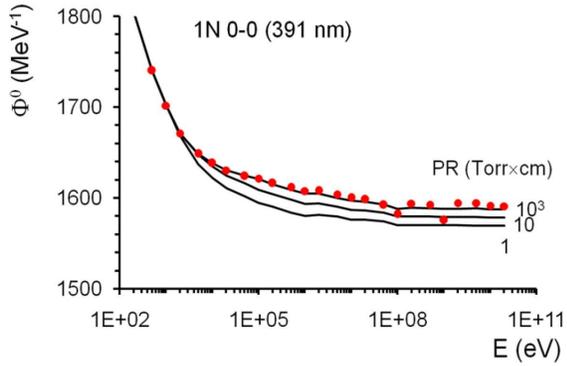}
\caption{\label {fig7} Same as figure 6 for the 1N (0-0) transition.}
\end{center}
\end{figure}

\par

These figures show several interesting features. In the first place, for high energy (beyond the keV range), and for observation
regions larger than about 1 Torr$\times$cm, $\Phi^0$ is nearly constant. Nevertheless a smooth dependence on primary energy is
found. The value of $\Phi^0$ decreases with primary energy about 10\% in the range 1 keV - 1 MeV and 4\% in the interval 1 MeV -
20 GeV for the 337 nm band (figure 6). For the 391 band the corresponding decreasing is about 6\% for the interval 1 keV - 1 MeV
and 1\% for 1 MeV - 20 GeV (figure 7). On the other hand a smooth growing of $\Phi^0$ with $PR$, smaller than 2\% in the range
10 - 1000 Torr$\times$cm, is also found for energies larger than 1 MeV (figures 8 and 9). At lower energy and/or region size the
fluorescence is clearly not proportional to the deposited energy.
\par
The real number of photons detected per unit of deposited energy $\Phi$ can be calculated from equation (\ref{Phi1}) using the
corresponding $P'_v$ value. Unfortunately there are significant disagreements in the available experimental data of collisional
quenching in air. Therefore the comparison of the absolute value of the efficiency with experimental data is not
straightforward. On the other hand, some experiments only provide a measure of the total fluorescence in a wide spectral range
including most of the molecular transitions. In addition, very often the fluorescence yield is given in units of photons per
meter.

\begin{figure}[t]
\begin{center}
\includegraphics*[width=0.47\textwidth,angle=0,clip]{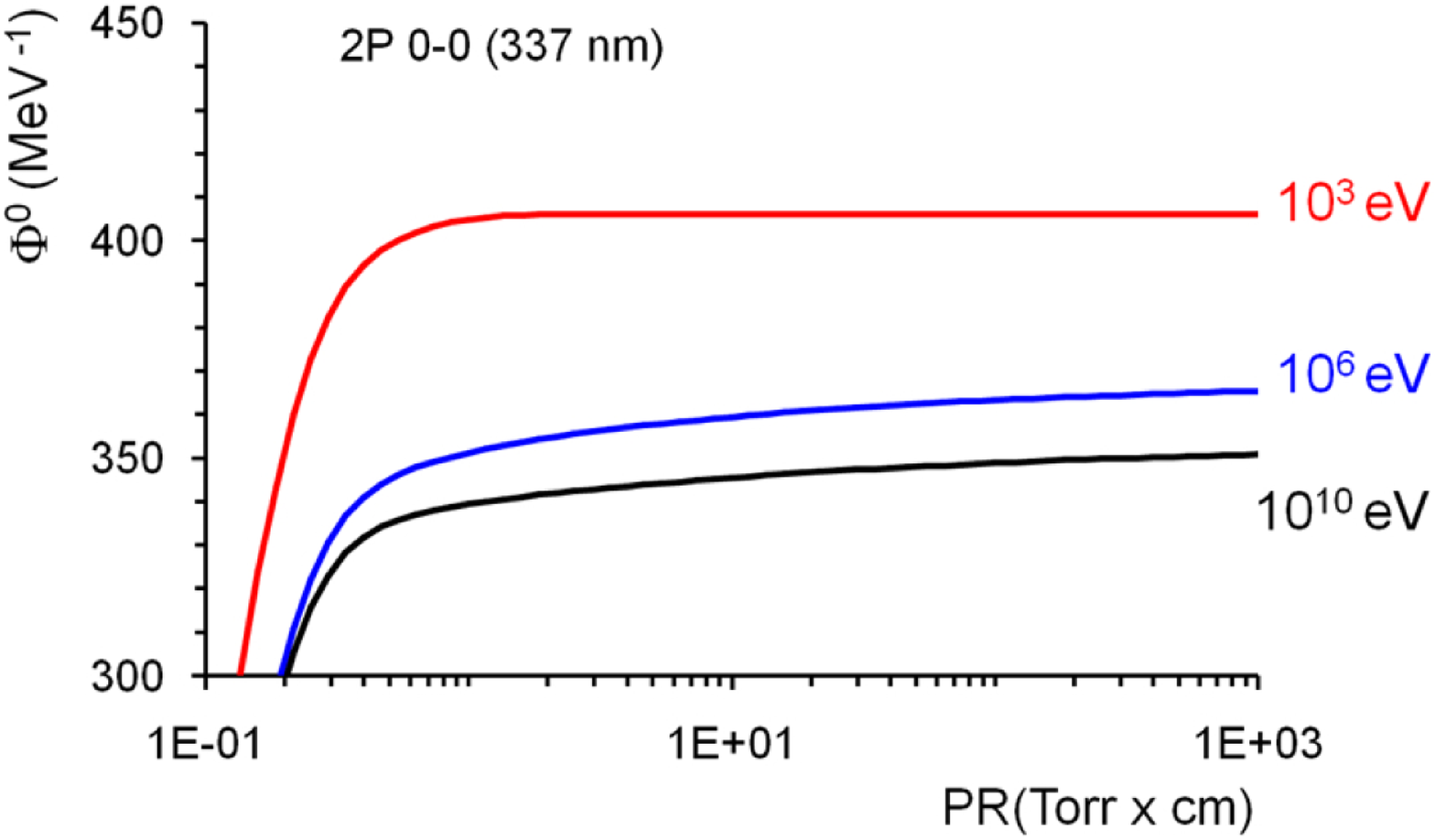}
\caption{\label {fig8} Fluorescence efficiency for the 2P (0-0) band in the absence of collisional quenching against the size of
the observation region $PR$ for several primary energies.}
\end{center}
\end{figure}

\begin{figure}[t]
\begin{center}
\includegraphics*[width=0.47\textwidth,angle=0,clip]{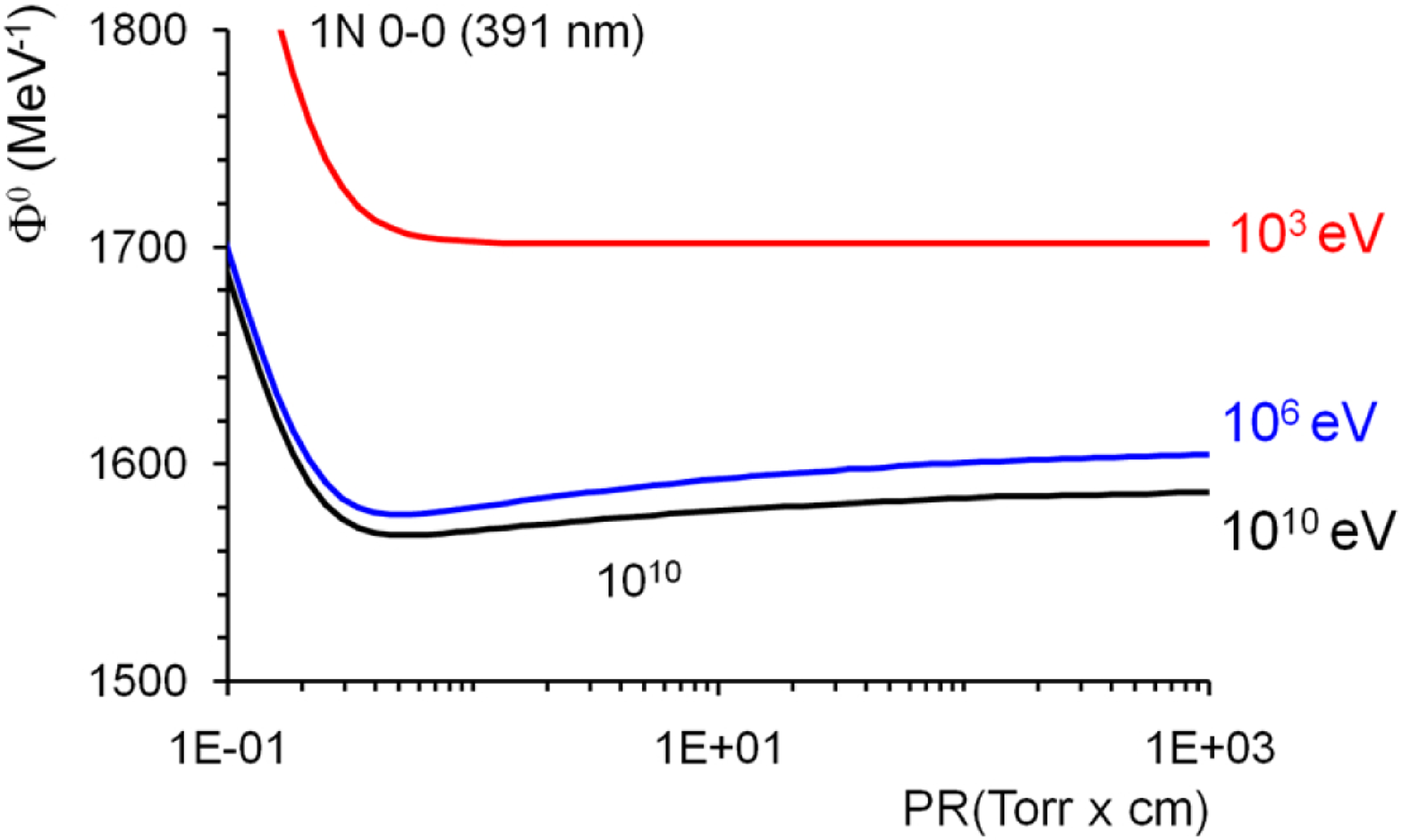}
\caption{\label {fig9} Same as figure 8 for the 1N (0-0) transition.}
\end{center}
\end{figure}

\par

The fluorescence efficiency at zero pressure for a reference molecular band (e.g. the 337 nm band) can be inferred from the
number of photons per unit electron path length $\varepsilon_{\lambda}$ emitted in a given wavelength interval at a given
pressure by

\small
\begin{equation}
\label{rel} \Phi^0_{337} = \frac{\varepsilon_{\lambda}}{{\rm d}E_{dep}/{\rm d}x}
\frac{I_{337}}{I_{\lambda}}\left(1+\frac{P}{P'_{337}}\right)\,,
\end{equation}
\normalsize
\small
\begin{table*}[!t]
\caption{Parameters for the calculation of $\Phi^0_{337}$ experimental data at atmospheric pressure. Experiment reference,
electron energy, spectral bandwidths and measured number of air-fluorescence photons per meter are displayed in columns one to
four. Fifth column shows the deposited energy per meter obtained from our simulations. The sixth one gives the total
fluorescence efficiency at atmospheric pressure in the spectral interval. The ratio of the 337 nm band intensity to total
intensity is shown in next two columns (7 and 8). Ratios have been calculated assuming the set of $P'_v$ values given either by
references \cite{Pancheshnyi} ($P'_P$ column) or \cite{airfly1} ($P'_A$ column). Next two columns show inferred values of
$\Phi^0_{337}$ assuming the above mentioned $P'_v$ values. Column eleven shows the $\Phi_{337}^0$ result reported by the
experiment. Last column shows the preliminary theoretical results presented in this work. Bolded figures are those inferred from
our calculations.}
\centering
\begin{tabular}{|lccccc|cc|cc|c|c|}
\hline

Experiment  &  ~~~~~~ $E$  ~~~~~~ &  ~~~~~~~ $\lambda$  ~~~~~~~ & ~~ $\varepsilon_\lambda$ ~~ &  ${\rm d}E_{dep}/{\rm d}x$  & ~
$\Phi_{\lambda}$~~
&\multicolumn{2}{c}{$I_{337}/I_{\lambda}$}&\multicolumn{4}{|c|}{$\Phi_{337}^0$[MeV]$^{-1}$}\\
~  &  [MeV]  &  [nm] & [m$^{-1}$] & [MeVm$^{-1}$] & ~~[MeV]$^{-1}$~~ & ~~$P'_P$ ~~ & ~~ $P'_A$ ~~ &  ~~$P'_P$ ~  & ~ $P'_A$ ~~ &
reported
& this work\\

\hline \hline

Nagano et al.      & 0.85                      & 337       & 1.021 & \textbf{0.18}  & \textbf{5.67} &  1   &  1    &  \textbf{440}  & \textbf{370}  & 272 & \textbf{365} \\
\hline
FLASH 07    & 2.8$\times$10$^4$         & 300 - 420 & 5.06  & \textbf{0.24}  & \textbf{20.8}$^\dag$ &  \textbf{0.28} &  \textbf{0.27} & \textbf{460}   & \textbf{370}  &  ~  & \textbf{350} \\
\hline
MACFLY      & 1.5 -  5.0$\times$10$^4$  & 290 - 440 & ~     & ~    &  17.6 & \textbf{0.27}  &  \textbf{0.26}  & \textbf{380}  & \textbf{290}  & 174  & \textbf{365 - 352}\\
\hline
\multicolumn{12}{l}{$^\dag$ Value reported by FLASH.}\\
\end{tabular}
\end{table*}
\normalsize where the ratio between the fluorescence intensity for the 337 nm band $I_{337}$ and that for a broad wavelength
interval $I_{\lambda}$ can be calculated from the Franck-Condon factors, Einstein coefficients and $P'_v$ values, as explained
in detailed in ref. \cite{arqueros}. For the comparisons shown below the energy deposited per unit path length ${\rm
d}E_{dep}/{\rm d}x$ has been obtained from our MC simulations (figure 5) assuming for all experiments a typical $PR$ value of
1600 Torr$\times$cm. Notice that this parameter depends on $PR$ very smoothly (around 6\% variation in the interval 300 - 3000
Torr$\times$cm). For the following discussion we will use $P'_v$ measurements very recently published by the AIRFLY
collaboration \cite{airfly1} and those obtained from deactivation rate constants N-N and N-O measured by Pancheshnyi et al.
\cite{Pancheshnyi} (see also \cite{arqueros}).

\par

We have compared our theoretical $\Phi^0_{337}$ value with some available experimental data. For instance Nagano et al.
\cite{nagano2}, working at an average energy of about 0.85 MeV, reports a value of 272 MeV$^{-1}$ significantly smaller than our
predictions. However from their measurement of 1.021 photons (337 nm) per meter at atmospheric pressure, results from
(\ref{rel}) of 440 or 370 MeV$^{-1}$ are found by using $P'_v$ values of \cite{Pancheshnyi} or \cite{airfly1} respectively, the
last one in very good agreement with our predicted value of 365 MeV$^{-1}$ (see table 1). This discrepancy can be explained by
comparing the $P'_{337}$ value of 14.4 Torr reported by \cite{nagano2} with those of 9.8 and 11.9 Torr from \cite{Pancheshnyi}
and \cite{airfly1} respectively. This simple exercise shows that the result of the comparison is strongly dependent on the
chosen $P'_{337}$ value.

\par

Another interesting example is the comparison with the measurement recently reported by the FLASH collaboration at 28 GeV
\cite{flash2}. In this case we use their 5.06 photons/m for the whole fluorescence spectrum in the wavelength range 300 - 420
nm. The contribution to the total intensity $I_{\lambda}$ of the Gaydon-Herman bands measured by AIRFLY \cite{airfly1} has been
included.  The inferred $\Phi^0_{337}$ results are 453 and 353 MeV$^{-1}$ using $P'_{337}$ values of \cite{Pancheshnyi} or
\cite{airfly1} respectively (see table 1). Again the second one is in very good agreement with our predictions. Note that, as
shown in the table, the factor $I_{337}/I_{\lambda}$ is only weakly dependent on the $P'_v$ set used.

\par

Finally the $\Phi^0_{337}$ result reported by MACFLY \cite{colin} of 174 MeV$^{-1}$, constant in the range 1.5 MeV - 50 GeV, is
smaller than our predictions. Notice that the value $P'_{337}$ = 19.4 Torr of \cite{colin} is larger than those of
\cite{Pancheshnyi} or \cite{airfly1}. In fact, following the above described procedure a $\Phi^0_{337}$ value of 380 or 290
MeV$^{-1}$, using \cite{Pancheshnyi} or \cite{airfly1} respectively, would be inferred from the MACFLY result of
$\Phi_{\lambda}$ = 17.6 photons/MeV at atmospheric pressure in the spectral interval 290 - 440 nm.

\par

\section{Conclusions}

An improved model for the air-fluorescence emission has been used to analyze in detail the relationship between fluorescence
intensity and deposited energy. For primary energies larger than a few keV and assuming observation regions over 10
Torr$\times$cm nitrogen fluorescence is nearly proportional to deposited energy. A smooth decrease of $\Phi^0$ of about 10\% in
the range 1 keV - 1 MeV and 4\% in the interval 1 MeV - 10 GeV for the 2P system is found. Taking into account that the
contribution of the energy released by an extensive air shower by electrons with energy smaller than 1 MeV is only of about 22\%
\cite{risse}, the above smooth dependence of efficiency on primary energy has no impact on the calibration of fluorescence
telescopes. Nevertheless very precise laboratory measurements (better than 10\%) of the fluorescence yield in the range keVs -
GeVs could observe this small effect.

\par

The absolute value of the fluorescence efficiency for the 337 nm band computed in this work has been compared with some
experimental results. The comparison was carried out by reducing experimental values of number of photons per meter at
atmospheric pressure to $\Phi^0_{337}$ using quenching data from other authors. In general we found a good agreement, in
particular when using quenching parameters from AIRFLY.

\par

Our improved energy distribution of secondaries (consistent with the energy loss predicted by the Bethe-Bloch formula) has lead
us to demonstrate quantitatively that fluorescence intensity is basically proportional to deposited energy. On the other hand,
uncertainties in the absolute values of theoretically predicted $\Phi^0_{vv'}$ quantities are larger because these are limited
by the accuracy of the assumed molecular parameters involved in the calculations.

\section{Acknowledgements}
\label{acknowledgements}

This work has been supported by the Spanish Ministry of Science and Education MEC (Ref.: FPA2006-12184-C02-01) and ``Comunidad
de Madrid" (Ref.: 910600).  J. Rosado acknowledges a PhD grant from ``Universidad Complutense de Madrid''.




\begin{thebibliography}{00}
\bibitem{EEDL}
EEDL databases are included in GEANT4 packages or available at www-nds.iaea.org/epdl97/.
\end{thebibliography}
\end{document}